# A Closed-loop Brain-Machine Interface SoC Featuring a 0.2μJ/class Multiplexer Based Neural Network

Chao Zhang[1], Yongxiang Guo[1], Dawid Sheng[1], Zhixiong Ma[2], Chao Sun[3], Yuwei Zhang[3], Wenxin Zhao[1], Fenyan Zhang[2], Tongfei Wang[2], Xing Sheng[1], Milin Zhang[1]

[1]Tsinghua University, [2]Chinese Institute for Brain Research, Beijing, [3]Beijing Ningju Technology

Closed-loop bidirectional brain-machine interface (CL-BBMI) consisting of a neural signal recorder, a neural signal processing unit (NPU) and a modulation module plays an important role in the cutting-edge research in neuroscience [1-2]. In addition to traditional electrical-mode modulation, optogenetic stimulation shows great potential in BBMI in recent years [3-4]. Among the three modules, the NPU is a key component for the closed-loop, but very limited solutions have been reported in literature [4]. A neural network (NN) is typically a high-accuracy solution for NPU but suffers unaffordable power consumption especially in the implantable scenarios of BMI. Table lookup based multiplier-free NN is a feasible low-power approach. However, the current lookup table (LUT) based table lookup solution still requires a huge quantity of random-access memory (RAM) [5].

This work presents the first fabricated electrophysiology-optogenetic CL-BBMI system-on-chip (SoC) with electrical neural signal recording, on-chip sleep staging and optogenetic stimulation. The first multiplexer with static assignment based table lookup solution (MUXnet) for multiplier-free NN processor was proposed. A state-of-the-art average accuracy of 82.4% was achieved with an energy consumption of only 0.2μJ/class in sleep staging task.

Fig. 1 (top) shows the block-diagram of the proposed CL-BBMI SoC, including an 8-channel low-noise amplifier (LNA) for neural signal acquisition, a cascaded integrator-comb (CIC) decimation filter for data compression, a MUXnet for neural signal processing and a dual-channel pulse width modulation (PWM) module for optogenetics. The memory unit of the MUXnet is divided into 6 small blocks, where a dynamic power gating is recruited to save 56.7% power consumption. An implantable, wireless optogenetic probe utilizing a blue color, thin-film micro-LED was integrated to interrogate the desired tissue according to the recognized sleep stage.

The inner product is the fundamental component of NNs. Given a bit-wise inner product $y^i=wx^i$ with a vector length of n and an output width of m (Fig. 1, bottom), a LUT based solution requires $m2^n+mn$ bits of memory for each table lookup when the quantization error is aligned to the MUXnet. In order to avoid the expensive memory access, the MUXnet selects the inner product results from the static VDD/GND assignment by MUXs, requiring only mn bits of weight memory to save the line index of the static assignment.

The process engine (PE) of the proposed MUXnet consists of a two-stage MUX based processing unit (MPU) architecture and an adder tree (Fig. 2, top right). The computing operation of the two-stage MPU relies on a static table (ST) (Fig. 2, top left) with two fundamental parameters n and m. The keys of the ST are all the possible conditions of the input activations. Each value line of the ST represents the inner product results between all the keys and one network weight vector with a length of n. In the stage-1 MPU, four MUXs are used for the weight selection from the ST. The trained weights are translated into the ST index in offline. The ST is statically assigned to VDD or GND. Although the ST itself is non-configurable according to the static assignment, the proposed MUXnet still supports arbitrary weight vectors in an NN by enumerating all the possible inner product results on chip. As shown in Fig. 2 (middle), different from the complete tables with an explosive number, the NN inference only requires a subset, denoted as inner product compatible (IPC) tables. The number of IPC tables is proportional to $2^{nm}$. It is possible to be enumerated under small n and m. As a result, only the ST index is required to be stored in the weight memory. An off-line pre-scaled weight scaling (Fig. 2, bottom) could be performed to reduce the quantization error without a requirement of more memory consumption, achieving an accuracy improvement of 61.4%.

As shown in Fig. 3 (top left), once a specific line of the ST is selected by the stage-1 MPU, the stage-2 MPU will produce the results of inner product with the input activations as keys. A bit-serial scheme is utilized to ensure the stage-2 MPU always takes a single bit of the input vectors. As a result, a MUX size of $2^n$ to 1 is enough for the stage-2 MPU. Multiple m-bit results are then accumulated through a parallel post-lookup merging unit (PLMU), featuring a capacity of 8 groups of 8x8 inner products in one clock cycle using 32 MUXs in the stage-2 MPU. Fig. 3 (bottom) examples how to realize 2 groups of 8x8 inner products using stage-2 MPU, where each n multipliers and n-1 adders are replaced by one $2^n$ to 1 MUX and a PLMU. The MUX based inner product features a 0.34x lower power consumption than a typical multiplier based inner product. Fig. 3 (top right) shows the MUXnet actually utilizes a non-linear quantization, holding a 0.69x smaller quantization interval to limit the computation error.

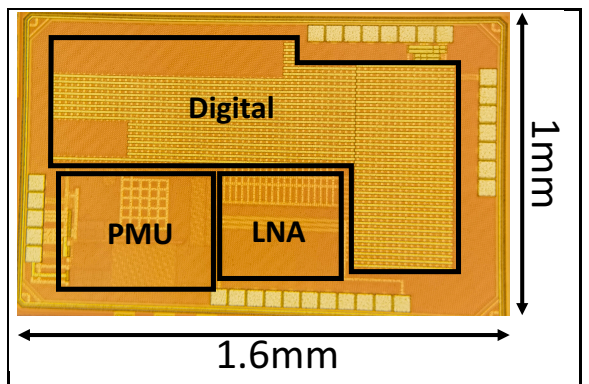

Die micrograph.

Fig. 4 (top left) illustrates the designed 4-layer CNN for sleep staging. The 1-d batch normalization (BN) layers are absorbed into the 1-d convolutional layers to free the weight memory and the on-chip inference for BN layers. Each optogenetic channel can be triggered after arbitrary predicts (0-9) produced by the MUXnet. The segment length for each classification is reconfigurable. The predictions of six 5-second segments would vote out the stage of one 30-second segment in the human sleep staging task. A class-wise early stop voting where the different predicts hold different thresholds to be the majority is proposed to reduce the classification times by up to 46.7% (Fig. 4, bottom left). A dual mode MUXnet with an m of 5 or 10 and an n of 2 was implemented. The convolutional layers utilize an m of 10 to achieve a lower quantization error, while the linear layers utilize an m of 5 to reduce the memory consumption. The stage-1 MPU with an m of 10 is decomposed by two small MPUs with m of 5 (Fig. 4, right). Since the table size grows exponentially with m, the table decomposition reduces the power consumption and area of stage-1 MPU with an n of 2 and an m of 10 by 59x and 548x, respectively.

The proposed CL-BBMI was fabricated in standard 40nm CMOS technology, occupying a silicon area of $1.6mm^2$. When verified on a public 20-subject database (DREAMS) and a private mice dataset, an average sleep staging accuracy of 82.4% and 84.3% was achieved, respectively (Fig. 5, top). Under a supply voltage of 0.73V and a system clock of 23MHz, an average power dissipation of 172.4μW and an energy consumption of 0.2μJ/class were achieved (Fig. 5, bottom right). The proposed SoC was then integrated with a RF chip for the wireless in-vivo verification. An 488nm optogenetic modulation with 10Hz frequency and 10% duty cycle was triggered once a non-rapid eye movement (NREM) phase was detected. Remarkable changes were detected from the electroencephalogram (EEG) and the electromyography (EMG) signals after the optogenetic stimulation (Fig. 5, bottom left), ensuring the effectiveness of the proposed CL-BBMI in cutting-edge neuroscience research.

Fig. 6 compares the proposed work with state-of-the-art designs. Compared with other optogenetics based BBMI SoCs [3-4], the proposed work highlights in the NN based complex closed-loop control. In the sleep staging task, this work shows the highest average accuracy and a 53x lower energy consumption per classification than the prior arts [6-7]. The MUXnet provides a new approach for table lookup based multiplier-free NN inference without the use of LUT. Featuring a local automatic closed-loop, this work offers a state-of-the-art powerful tool in frontier neuroscience study.

**Acknowledgement:**

Placeholder for acknowledgements.

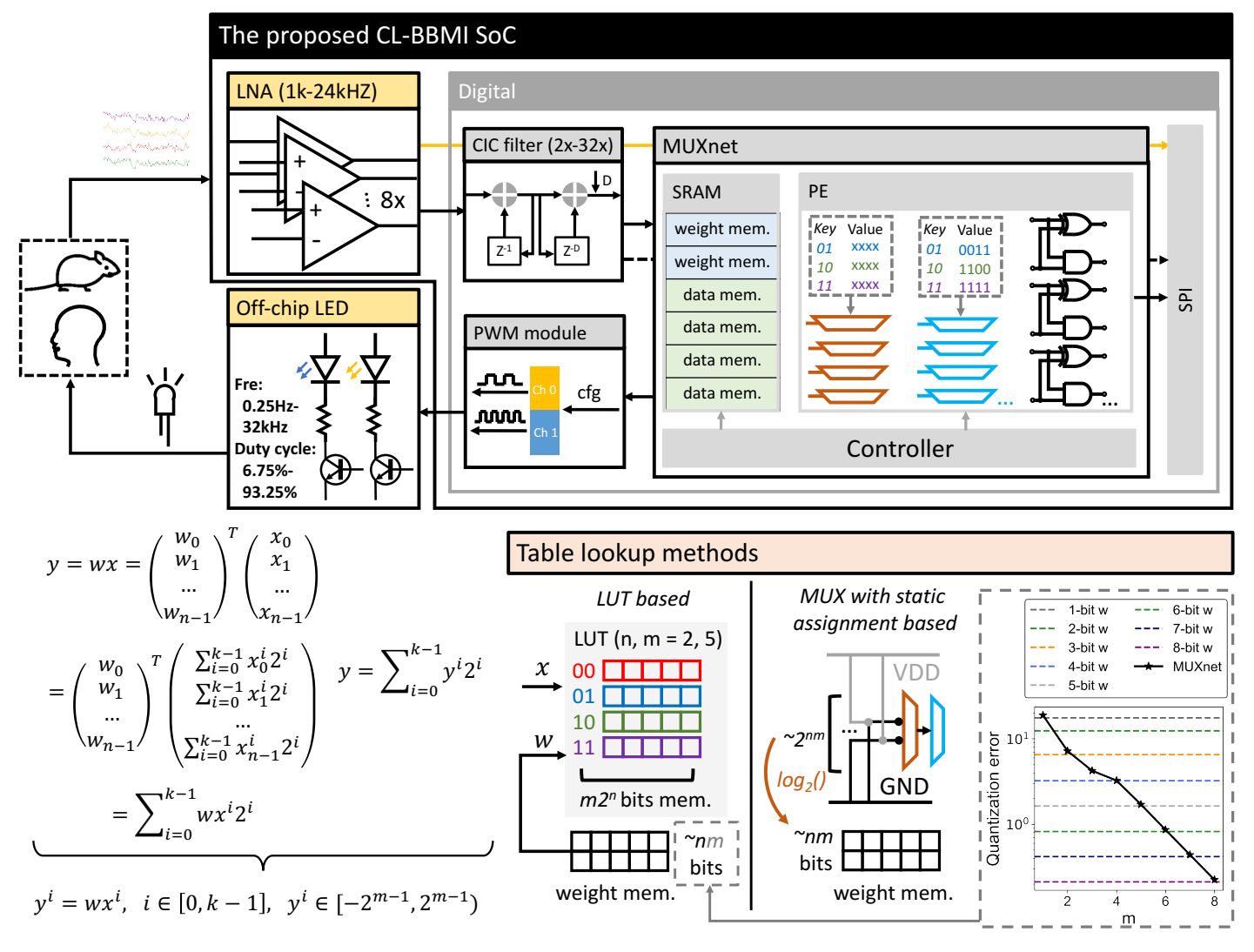

Fig. 1. (Top) The block-diagram of the proposed CL-BBMI. (Bottom) The table lookup methods for bit-wise inner product.

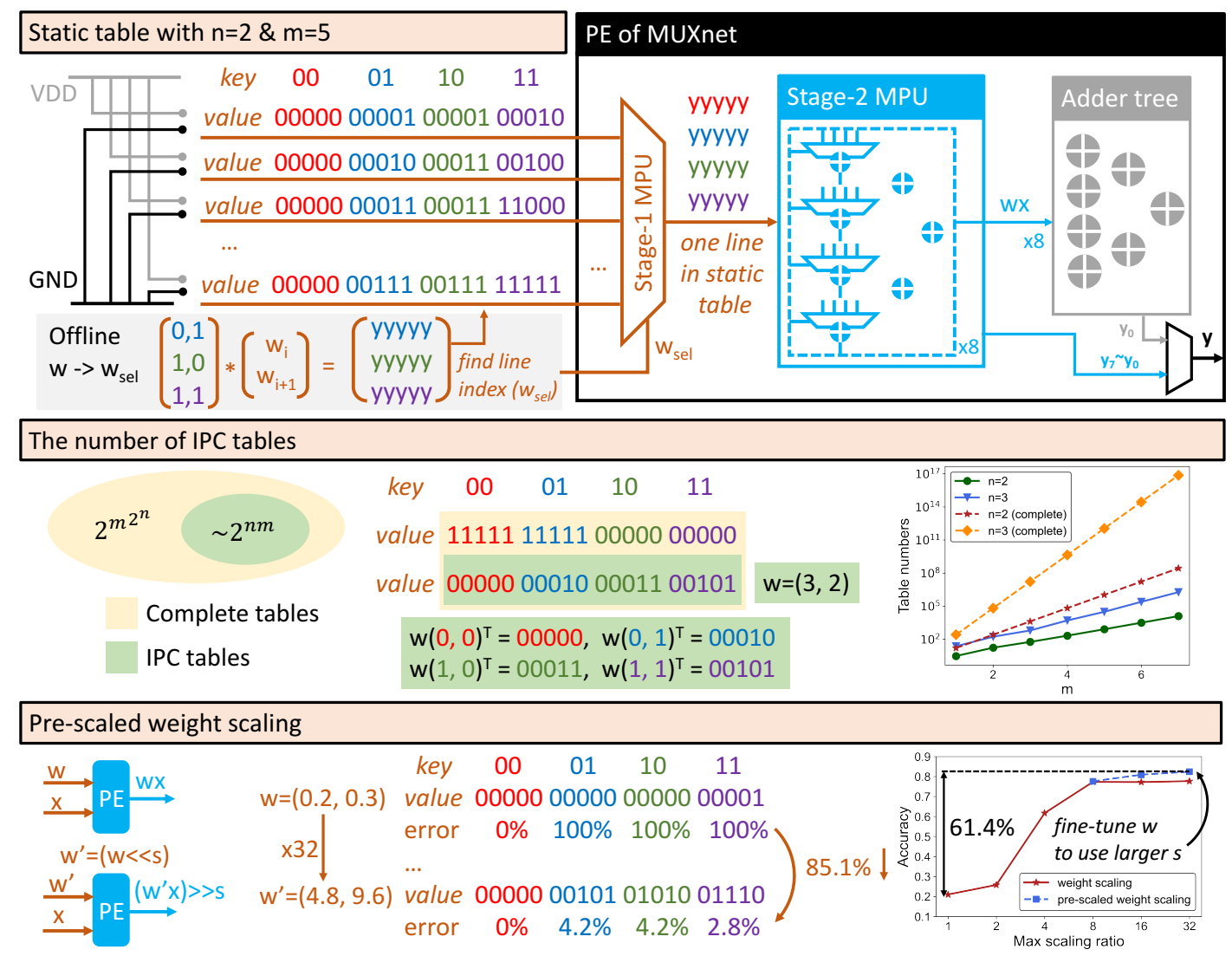

Fig. 2. (Top) The static table and the PE of MUXnet. (Middle) The number of IPC tables. (Bottom) The pre-scaled weight scaling.

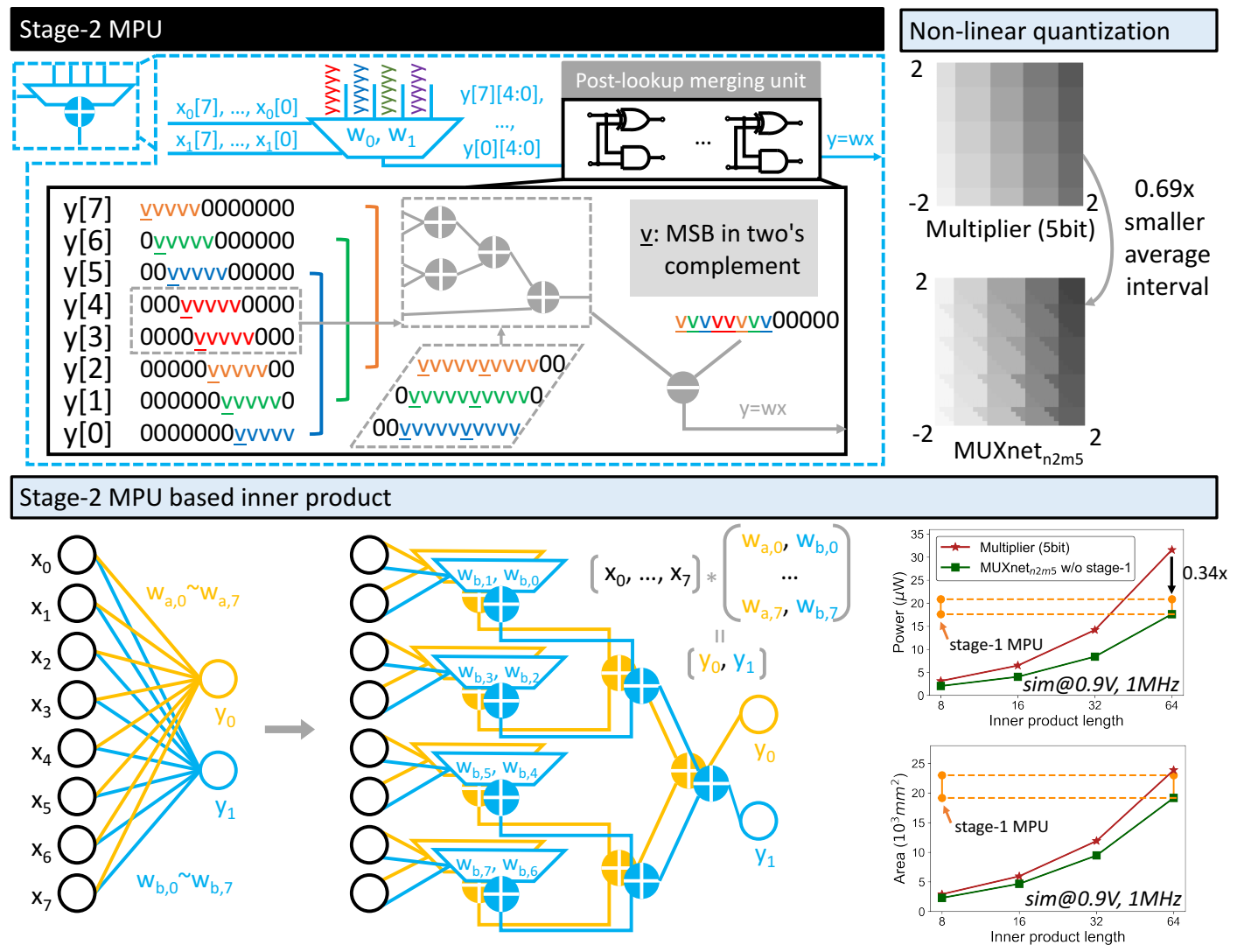

Fig. 3. (Top) The stage-2 MPU and the non-linear quantization in the MUXnet. (Bottom) The stage-2 MPU based inner product.

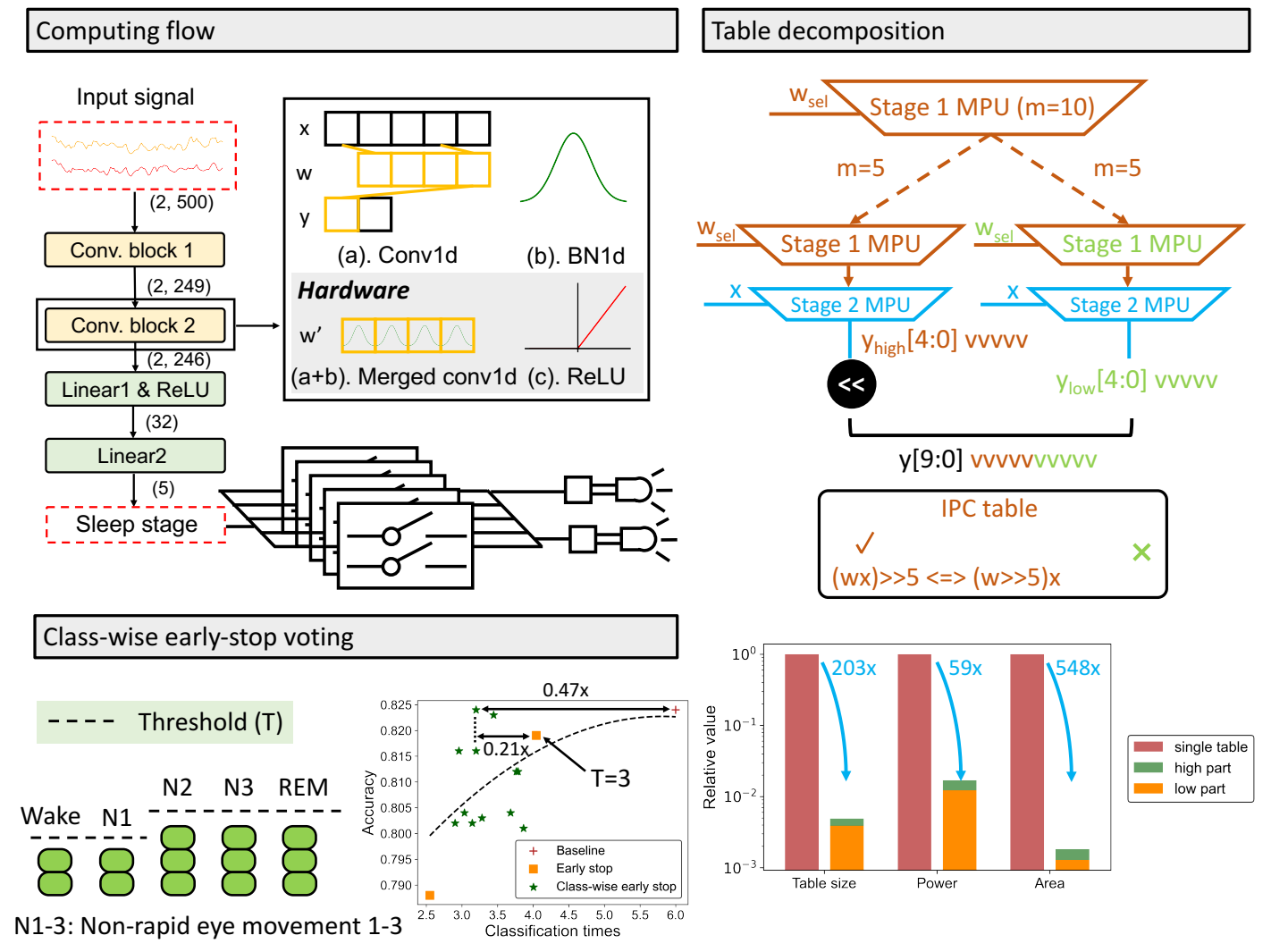

Fig. 4. (Top left) The computing flow. (Bottom left) The class-wise early-stop voting. (Right) The table decomposition.

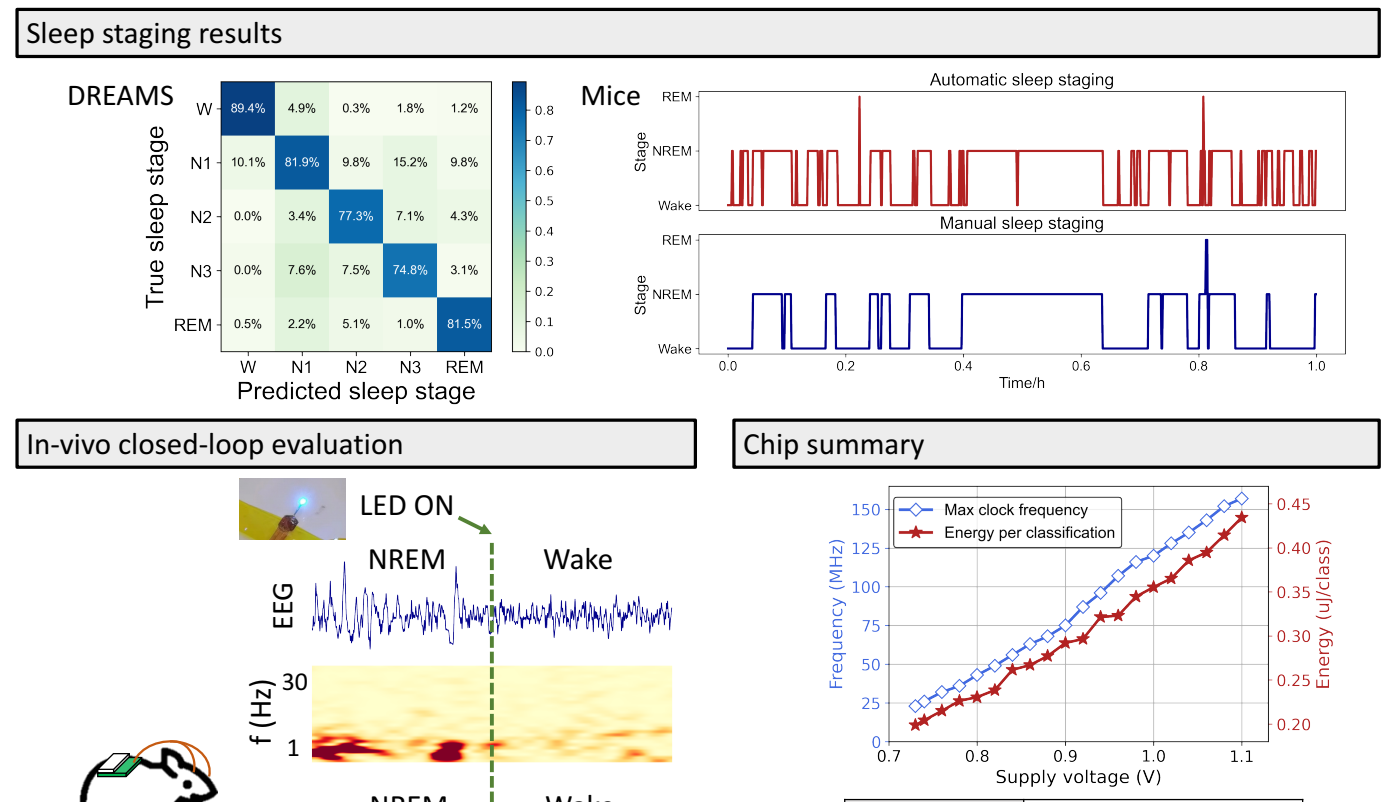

| Technology | 40nm CMOS | |
|---|---|---|
| Area | Chip | 1.60mm2 |
| | Digital | 0.68mm2 |
| Frequency | 23MHz-157MHz | |
| Supply | Core: | 0.73V-1.1V |
| | IO: | 1.8V/3.3V |
| Power | 172.4µW-2.57mW | |
| SRAM | Weight | 2048x40bit |
| | Data | 512x64bit |
| Bit precision | INT8 | |

Fig. 5. The test results.

| | ISSCC 18 [3] | ISSCC 19 [4] | JSSC 17 [6] | TCAS-I 19 [7] | This work |
|---|---|---|---|---|---|
| Technology | 130nm | 180nm | 180nm | 180nm | **40nm** |
| Area ($mm^2$) | 1.2 | 13.67 | 10.3 | 11.74 | **1.6** |
| NPU | | | | | |
| Algorithm | Filter | FE + LLS | FE + DT | FE + NN-based DT | **End-to-end CNN** |
| Application | - | Seizure detection | Sleep staging | Sleep staging | **Sleep staging** |
| Dataset | - | - | DREAMS | SHHS, NTUH | **DREAMS** |
| #Channel | - | 2 | 1 | 1~2 | **2** |
| Accuracy | - | - | <80%[a] | 77.1~81.1% | **82.4%** |
| Data width | - | - | 24b fixed | - | **8bit** |
| Supply (V) | - | 1.8 | 1.25 | 1.2 | **0.73** |
| Power (µW) | - | 114 @375Hz | 426 @1.5kHz | 4.96 @10kHz | **172.4 @23MHz** |
| Energy (µJ/class)[b] | - | - | 10.5 | 149 | **0.2** |
| On-chip SRAM | None | None | None | None | **14kB** |
| Recorder & Stimulator | | | | | |
| LNA channel | 10 | 2 | - | - | **8** |
| Stim channel | 4 | 1 | - | - | **2** |
| Closed-loop | N | LLS based | - | - | **NN based** |
| Stim type | Optogenetic | Optogenetic | - | - | **Optogenetic** |
| Device weight | - | 6g | - | - | **3.8g[c]** |
| In-vivo validation | Y | Y | - | - | **Y** |

Bi-directional brain-machine interface SoCs — Sleep-staging ASICs a: The reported 98.5% is the consistency between hardware and software.

b: Thirty second per classification.

c: Including a 2.5g battery.

| | Nature 2015 [1] | Nat. NeuroSci. 2022 [2] | This work |
|---|---|---|---|
| Automatic closed-loop | Yes | Yes | **Yes** |
| Local classification | No | No | **Yes** |

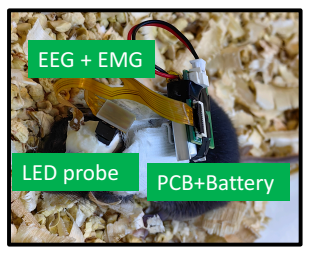

Fig. 6. The comparison tables and the setup of in-vivo test.